\documentclass[11pt, a4paper]{scrartcl}
\usepackage[USenglish]{babel} 
\usepackage{graphicx} 
\usepackage{amsmath} 
\usepackage{amssymb} 
\usepackage[
thmmarks, 
amsmath, 
hyperref 
]{ntheorem} 
\usepackage[hang]{subfigure} 
\usepackage{wrapfig} 
\usepackage{tabularx} 
\usepackage{booktabs} 
\usepackage{listings} 

\newtheorem{theorem}{Theorem}
\newtheorem{remark}{Remark}
\theoremsymbol{$\Box$}

\newcommand*{\IP}{\mathbb{P}}
\newcommand*{\IE}{\mathbb{E}}
\newcommand*{\IR}{\mathbb{R}}
\newcommand*{\IN}{\mathbb{N}}

\newcommand*{\F}{\mathcal{F}}

\begin{document}
\nocite{*}

	\begin{center}
		{\bfseries\Large Pricing bonds with optional sinking feature using Markov Decision Processes}
		\par\bigskip
		\vspace{1cm}
		
		{\Large Jan-Frederik Mai}\\
		\vspace{0.2cm}
		{XAIA Investment GmbH,}\\
		{Sonnenstra{\ss}e 19, 80331 M\"unchen, Germany,}\\
		{email: jan-frederik.mai@xaia.com,}\\
		{phone: +49 89 519966-131.}\\
		\vspace{0.8cm}
		{\Large Marc Wittlinger}\\
		\vspace{0.2cm}
		{Institute of Mathematical Finance, Ulm University,}\\
		{Helmholtzstra{\ss}e 18, 89081 Ulm, Germany,}\\
		{email: marc.wittlinger@uni-ulm.de,}\\
		{phone: +49 731 50-23557.}\\
	\end{center}

\vspace*{1.2cm}

An efficient method to price bonds with optional sinking feature is presented. Such instruments equip their issuer with the option (but not the obligation) to redeem parts of the notional prior to maturity, therefore the future cash flows are random. In a one-factor model for the issuer's default intensity we show that the pricing algorithm can be formulated as a Markov Decision Process, which is both accurate and quick. The method is demonstrated using a $1.5$-factor credit-equity model which defines the default intensity in a reciprocal relationship to the issuer's stock price process, termed jump-to-default extended model with constant elasticity of variance (JDCEV) in \cite{carr06}.

\vspace*{0.7cm}

\section{Introduction}
In the high-yield credit market it is not uncommon that bond issues come equipped with a callable feature. This allows the issuer to pay back the bond's notional prior to maturity, possibly with an additional fee. Such features complicate the mathematical treatment massively compared with plain vanilla instruments, since the future cash flows generated by the bond are random and depend on the issuer's decisions in the future. Resorting to backwardation techniques known from equity derivative pricing, callable bonds can be evaluated in standard diffusion models for the short rate and/or the issuer's default intensity. The idea is to build an approximating tree for the underlying driving process(es). Working backwards in time, the issuer has to decide at each time point whether she wants to make use of her option and call the bond or not. Hence, the value of the callable bond at each node in the tree is determined as the minimum of  the call strike and the continuation value at the respective
time point. Similar algorithms apply to bonds with put and/or conversion  features.
\par
However, the picture becomes dramatically more difficult when the issuer is not only allowed to decide when she redeems the notional, but additionally can distribute her total redemption into several installments of optional size. Such bonds are very unusual, however, they exist. For example, Westvaco Corporation, a US packaging company, has issued a $150\$$ billion bond with optional sinking feature in March 1997, whose ultimate maturity is June 2027. On an annual basis the issuing company is allowed to redeem either $5\%$ or $10\%$ of the outstanding notional. The aforementioned tree pricing approaches cannot be applied directly to such a situation because at each node in the tree, one does not only have to make a dichotomic decision between "redemption" or "continuation", but instead has to make a decision about the size of the redemption schedule.  In the present article, we show how such an additional feature can be incorporated into tree pricing approaches by formulating the problem as a Markov
Decision Process. The resulting algorithm is simple to implement and computationally efficient. As a special case our ansatz includes the case of a regular callable bond.


The remainder of the present article is organized as follows: Section \ref{sec_model} introduces the modeling assumptions we apply, Section \ref{sec_MDP} formulates the pricing algorithm as a Markov Decision Process, Section \ref{sec_ex} provides a numeric example, and Section \ref{sec_con} concludes.

\section{The model} \label{sec_model}
Generally speaking, the price of a fixed-income security issued by a company depends on the future evolution of interest rates (used for discounting cash flows), the future evolution of the market's opinion about the company's creditworthiness (typically encoded in the so-called default intensity, see below), and the random recovery payment to be received in the case of a default of the issuer during the instrument's lifetime. Ideally, a stochastic model for all three stochastic, and possibly dependent, factors is set up and used for the pricing. However, the need for practical viability requires one to build simpler models, especially when the considered instrument is equipped with American-style features which necessitate the use of tree methods. Even though this is possible theoretically, tree pricing algorithms with a satisfactory level of accurateness and computational speed typically exist only for one-factor models. Since we are especially concerned with fixed income derivatives in the high yield
sector the most dominant stochastic driver among the three sources of randomness is the default intensity. We therefore assume that the discounting interest rate curve, as well as the recovery rate are deterministic in order to end up with a one-factor model for the issuer's default intensity.
\par
Mathematically, we work on a probability space $(\Omega,\F,\IP)$ supporting a positive stochastic process $\lambda=\{\lambda_t\}_{t \geq 0}$ and an independent exponential random variable $\epsilon$ with unit mean. The stochastic process $\lambda$ is interpreted as the issuer's default intensity which is assumed to be observable in the marketplace. Technically, this is achieved by defining the issuer's default time $\tau$ via the so-called canonical construction
\begin{gather}
\tau = \inf\Big\{ t>0\,:\,\int_{0}^{t}\lambda_s\,ds>\epsilon\Big\}
\label{canonical}
\end{gather}
which yields the first order approximation $\IP(\tau \leq t+\Delta\,|\,\tau>t) \approx \lambda_t\,\Delta$, justifying the nomenclature "default intensity", see, e.g., \cite[Chapter 3]{jeanblanc06}. Defining the market filtration $\{\F_t\}_{t \geq 0}$ via $\F_t=\sigma(\lambda_s\,:\,s \leq t) \vee \sigma(1_{\{\tau>s\}}\,:\,s \leq t)$ implies that market participants can observe the default intensity and the default event, but $\tau$ is an unpredictable stopping time  (since $\epsilon$ cannot be observed). The assumption that $\lambda$ can be fully observed by the market is simplifying but not too unrealistic, because one typically can observe credit default swap spreads and other derivative prices related to the issuer on a daily basis, from which (total) information about $\lambda$ can be extracted. Given this setup, at time $t$ a zero coupon bond with maturity $T>t$ issued by the considered company has the market price
\begin{align}
1_{\{\tau>t\}}\,ZCB(t) &= 1_{\{\tau>t\}}\,\IE\Big[ 1_{\{\tau>T\}}\,e^{-\int_{t}^{T}r_s\,ds} + 1_{\{t<\tau \leq T\}}\, R\,e^{-\int_{t}^{\tau}r_s\,ds} \,\Big|\,\F_t\Big]
\label{001}
\end{align}
where $R \in [0,1]$ denotes the constant recovery rate and $\{r_t\}_{t \geq 0}$ denotes a discounting rate, which in the sequel will always be modeled in a deterministic manner.
\par
For the numerical pricing of American-style instruments it is necessary to discretize the model appropriately. To this end, we will approximate $\lambda$ by a discrete-time process $\{\lambda_{t_n}\}_{n \in \IN_0}$ on a time grid $0 =t_0<t_1<t_2<\ldots < t_N = T$ with finite state space -- a so-called tree approximation. To simplify notation, let us write
\begin{gather*}
\bar{F}_n:=e^{-\sum\limits_{i=1}^{n}\lambda_{t_{i-1}}\,\Delta t_i} \approx \IP\big(\tau>t_n\,\big|\,\{\lambda_t\}_{t \geq 0}\big),\quad \Delta t_i:=t_i-t_{i-1}.
\end{gather*}
Applying the discrete-time approximation to Equation (\ref{001}), and discretizing appearing integrals respectively, this yields
\begin{align*}
1_{\{\tau>t_n\}}\,ZCB(t_n)& \approx 1_{\{\tau>t_n\}}\,\IE\Bigg[ e^{-\sum\limits_{i=n+1}^{N}r_{t_{i-1}}\,\Delta t_i}\,\frac{\bar{F}_N}{\bar{F}_n}\,\Bigg|\,\F_{t_n}\Bigg]\\
& \qquad + 1_{\{\tau>t_n\}}\,\IE\Bigg[R\,\sum_{i=n+1}^{N} e^{-\sum\limits_{j=n+1}^{i}r_{t_{j-1}}\,\Delta t_j}\,\frac{ \bar{F}_{i-1}-\bar{F}_i}{\bar{F}_{n}}\,\Bigg|\,\F_{t_n}\Bigg]\\
& =: 1_{\{\tau>t_n\}}\,\big( \mbox{EDSC}(t_n)+\mbox{EDDC}(t_n)\big),
\end{align*}
where $EDSC$ stands for \underline{e}xpected \underline{d}iscounted \underline{s}urvival \underline{c}ashflows, and $EDDC$ for \underline{e}xpected \underline{d}iscounted \underline{d}efault \underline{c}ashflows. Since $\lambda_{t_n}$ is $\F_{t_n}$-measurable, one may take the terms $\bar{F}_{n+1}/\bar{F}_n$ out of the expectation values above to obtain

\begin{align}
&1_{\{\tau>t_n\}}\,ZCB(t_n) \nonumber \\
& \qquad = 1_{\{\tau>t_n\}}\,e^{-r_{t_n}\,\Delta t_{n+1}}\,\Bigg( R\,\Big( 1-\frac{\bar{F}_{n+1}}{\bar{F}_n}\Big)  + \frac{\bar{F}_{n+1}}{\bar{F}_n }\,\IE\big[EDSC(t_{n+1})+EDDC(t_{n+1})\,\big|\,\F_{t_n}\big] \Bigg) \nonumber\\
& \qquad = 1_{\{\tau>t_n\}}\,\underbrace{e^{-r_{t_n}\,\Delta t_{n+1}}}_{ \stackrel{\mbox{1-step}}{\mbox{discount}}}\,\Bigg( \underbrace{\Big( 1-\frac{\bar{F}_{n+1}}{\bar{F}_n}\Big)}_{ \stackrel{\mbox{1-step}}{\mbox{def. prob.}}}\,R +\underbrace{\frac{\bar{F}_{n+1}}{\bar{F}_n }}_{ \stackrel{\mbox{1-step}}{\mbox{surv. prob.}}}\,\IE\big[ZCB(t_{n+1})\,\big|\,\F_{t_n}\big] \Bigg)
\label{backwardationformula}
\end{align}
The last formula is the well-known recursion for $ZCB(t_n)$ which, starting from $ZCB(t_{N})=1$, allows to derive the desired value $ZCB(t_0)$ via backwardation techniques, where the conditional equations in each step are computed within the tree making use of the tower property of conditional expectation.
\par
There are many models for $\{\lambda_t\}_{t \geq 0}$ which allow to derive a closed formula for the value $ZCB(t_0)$ of a zero coupon bond, and hence by linearity for arbitrary coupon bonds. For instance, this is the case in the JDCEV model of \cite{carr06}, which we are going to employ below to demonstrate our approach. However, we are concerned with coupon bonds that are more exotic. Especially in the high-yield corporate credit markets it is the rule rather than the exception that bond issues come equipped with call rights for the issuer. We go even further by assuming the issuer has the right to redeem the notional in several installments of her choice. Such exotic features make numerical schemes like tree approximations and backwardation pricing techniques necessary, which is why we presented the general idea above for later reference.

\section{Pricing via Markov Decision Processes} \label{sec_MDP}
To evaluate a bond with optional sinking feature we formulate the pricing problem as a Markov Decision Process (short: MDP) with a finite horizon $N\in\IN$. For a detailed introduction and discussion of MDPs we refer to \cite{BR} or \cite{BS}.

From a given continuous-time model for the default intensity $\lambda = \{\lambda_t\}_{t\ge 0}$, we construct a tree approximation $\{\lambda_{t_n}\}_{n =0,1,\ldots,N}$ on a discrete-time grid $0=t_0<t_1<t_2<\ldots<t_N = T$. Moreover the random variable $\lambda_{t_n}$, called default intensity rate at time $t_n$, is assumed to take values in a finite set $Z^{(n)}$ whose elements are denoted\footnote{$|Z^{(n)}|$ denotes the cardinality of the finite set $Z^{(n)}$.} $Z^{(n)}:=\{z^{(n)}_i\}_{i =1,\ldots,|Z^{(n)}|}$. In the following $S\times Z$ will be the state space of the MDP where $S:=\{0,\tfrac{1}{K},...,\tfrac{K-1}{K},1\}$ for $K\in\IN$ describes the remaining nominal and $Z:=\cup_{n=0}^{N}Z^{(n)}$ the current value of the default intensity rate. To the mentioned state space we add an artificial cemetery state $\Theta$ which indicates a default of the bond. Furthermore, we  assume that the issuer gets once a fixed percentage $R\in [0,1]$ of the remaining nominal $s \in S$
when the bond defaults. Now we formulate the pricing problem.

\begin{center}
 {\bf \large Markov Decision Process with a finite horizon}
\end{center}
\begin{itemize}
\item State space $E:= S \times Z$ endowed with the Borel $\sigma$-algebra $\mathcal{E}$, where $s \in S$ is the remaining nominal and $z \in Z$ is the current default intensity.  Moreover, there exists a cemetery state $\Theta \notin E$.
\item Action space $A:= S$ endowed with the Borel $\sigma$-algebra $\mathcal{U}$.
\item The possible state-action combinations at time $n$ are given by
\begin{align*}
D_n&:=\{(s,z,a): (s,z)\in E\;, a\in D_n(s,z) \} \cup \{(\Theta,0)\}, \;0\le n< N\;,
\end{align*}
where the admissible actions in the state $(s,z)$ at time $n$ are specified by the set $D_n(s,z)$ satisfying
\begin{align*}
D_n(s,z) &\subset [0,s]\cap S\;,\quad D_{N-1}(s,z) = \{s\}\;,\quad   \forall (s,z)\in E\;,\; 0\le n<N-1\;,\\
D_n(\Theta) &= \{0\}\;,\quad 0\le n<N\;.
\end{align*}

Intuitively, depending on the remaining nominal $s$, the issuer is allowed to redeem any amount in $D_n(s,z)$ at time $t_n$. Since she can redeem at most $s$, $D_n(s,z)$ must be a subset of $[0,s]$.
\item The stochastic transition kernel $Q_n$ from $D_n$ to $E$ is given by
\begin{align*}
Q_n\Big[\Big(s-a,z^{(n+1)}_i \Big)\; \Big| \;  (s,z,a) \Big] := p^{(n)}_i(z),\quad 0\le n < N\;,
\end{align*}
where $i$ is running in the finite set $\{1,2,\ldots,|Z^{(n+1)}|\}$ and the probabilities $p^{(n)}_{i}(z)$ fulfill
\begin{align*}
\sum_{i =1}^{|Z^{(n+1)}|}p^{(n)}_{i}(z) =e^{-z\,\Delta t_{n+1}}\;.
\end{align*}
Further we set $Q_n \big[\Theta\; \big| \;  (s,z,a) \big] := 1-e^{-z\,\Delta t_{n+1}}\,$, $\,Q_n[\Theta\; | \;  (\Theta,0)] := 1$ and assign no probability mass to all remaining states of $E$. Intuitively, $e^{-z\,\Delta t_{n+1}}=\bar{F}_{n+1}/\bar{F}_n$ is the one-step survival probability at time $t_n$ when $z=\lambda_{t_n}$.
\item Denoting by $C_1,\ldots,C_N$ coupon rates we earn at time points $t_1,\ldots,t_N$, the one-stage cost function $c_n: D_n\rightarrow \IR $ is given by
\begin{align*}
c_n(s,z,a) &= \, e^{r_{t_n}\Delta_{t_{n+1}}}  \,\Big( (a + C_{n+1}\cdot s)\cdot e^{-z\,\Delta t_{n+1}}  + (1-e^{-z\,\Delta t_{n+1}})\, R \,s  \Big),\\
c_n(\Theta,0) &= 0, \quad 0\le n < N\;.
\end{align*}
If the issuer does not default until the next time point $t_{n+1}$
we earn the coupon payment $C_{n+1}\, s$ and the redemption amount $a$ the issuer has chosen. In the case of a default, we do not get these payments, but instead end up with the recovery fraction $R$ of the remaining nominal $s$. The following remark provides a comprehensive discussion of the cost functions. Regarding the coupons, notice that for most time points $t_n$ we have $C_n=0$, since the time grid is typically finer than the periodicity of the coupon payments.
\end{itemize}

\begin{remark}
The bond payments are illustrated in Figure \ref{fig-12} below for $N=3$. Thereby $a_n$ denotes the redemption amount which is paid at time $t_{n+1}$.
\begin{figure}[ht]
\center
\includegraphics[trim=3.2cm 24cm 0cm 2cm]{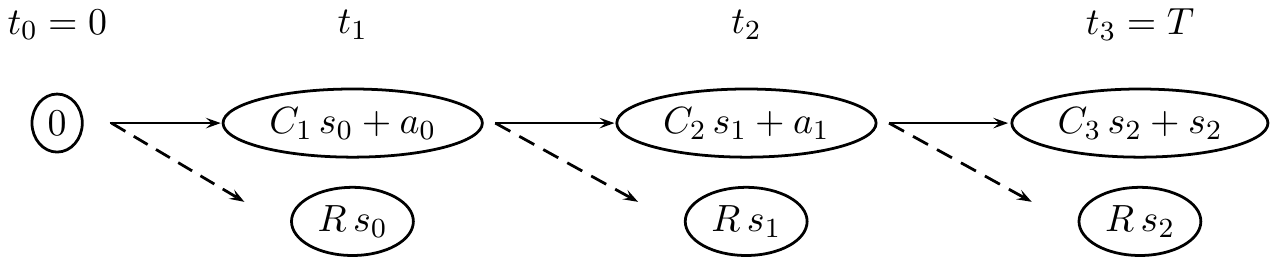}
\caption{Illustration of the bond payments at the time points $(t_n)_{0\le n\le 3}$. If the bond defaults then we go along the dashed arrow and get once the recovery payment.}
\label{fig-12}
\end{figure}
Shifting the payments one time-step back yields the cost function $\tilde{c}_n$ with
$$\tilde{c}_n(s_n,z_n,a_n,s_{n+1}) = \begin{cases} (C_{n+1}\,s_n+a_n)\,e^{-r_nt_{n+1}} &\mbox{, if } s_{n+1}\neq\Theta\;, \\
 R\, s_n\,e^{-r_nt_{n+1}} &\mbox{, if } s_{n+1}=\Theta\;, \end{cases}$$
for $0 \leq n<N$.  Since $\tilde{c}_n$ depends on the next (future) state we use the expected costs as proposed in \cite[Remark 2.1.2 b)]{BR} to dissolve this dependence. This yields the cost function $c_n$.
\label{283}
\end{remark}

The admissible action will be chosen by a decision rule, which is a measurable mapping $f:E\cup \{\Theta\} \rightarrow A$, such that $f(s,z)\in D(s,z)\;,\; \forall (s,z)\in E$ and $f(\Theta) =0$. Moreover, we define a Markovian policy $\pi$ as a sequence of decision rules, i.e. $$\pi:= (f_0,f_1,f_2,f_3,...,f_{N-1})\;,$$ where $f_k$ is a decision rule for each $k$. As usual in MDP theory, let $(\Omega,\mathcal{F})$ with $\Omega:=E^{N+1}$ and $\mathcal{F}:=\mathcal{E}\otimes ...\otimes \mathcal{E}$  be the measure space on which the state process $(S_n,Z_n)_{0\le n\le N}$ is defined. Then the $n$-th projection $(S_n,Z_n)(\omega)=(s,z)$ is the state of the MDP at time $n$. The theorem of Ionescu-Tulcea yields for a given Markovian policy $\pi$ that there exists a unique probability measure  $\IP^{\pi}_{(s_0,z_0)}$ on $(\Omega,\mathcal{F})$ such that
\begin{align*}
\IP^{\pi}_{(s_0,z_0)}\big{(}(S_{n+1},Z_{n+1})\in \,B\, | (S_{n},Z_{n})\big{)}= Q_n\big(\, B\, |(S_{n},Z_{n}),f_n(S_{n},Z_{n}) \big)
\end{align*}
and $\IP^{\pi}_{(s_0,z_0)}\big{(}(S_{0},Z_{0})\in B\big) = \delta_{(s_0,z_0)}(B)$ holds for all $B\in \mathcal{E}$ where $\delta_{(s_0,z_0)}$ denotes Dirac measure at $(s_0,z_0)$. Then $\IE^{\pi}_{n(s,z)}[\,.\,]$ denotes the expectation with respect to the probability measure $\IP^{\pi}_{n(s,z)}(\,.\,):=\IP^{\pi}_{(s_0,z_0)}(\,.\, | (S_n,Z_n) = (s,z))$. For $n=0,...,N$ and a Markovian policy $\pi$ we define the expected costs at time $n$ over the remaining states $n$ to $N$ and start $(s,z)\in E$ by
\begin{align*}
V_{n,\pi}(s,z) &:= \IE^{\pi}_{n(s,z)}\bigg{[} \sum_{k=n}^{N-1}  e^{-\sum\limits_{i=n+1}^{k}\hspace{-0.1cm}r_{t_{i-1}}\,\Delta t_i}\hspace{-0.3cm}c_k\big((S_k,Z_k),f_k(S_k,Z_k)\big)\bigg{]}  \;, \quad 0\le n\le N\;.
\end{align*}
Furthermore, we define the value function of the MDP by
\begin{align}
V_n(s,z) &:=  \inf_{\pi \in\Pi } V_{n,\pi}(s,z) , \quad  (s,z)\in E\;,
\label{1-006}
\end{align}
where $\Pi$ is the set of all Markovian policies.

Since the one-stage cost function is positive and we are minimizing the costs, the MDP is well defined, cp. \cite[Integrability Assumption p.17 and Remark 2.3.15]{BR}. Applying \cite[Theorem 2.3.8]{BR} yields the solution of the MDP, which is stated below in Theorem \ref{Bellman}. To formulate that solution in a convenient way we introduce the operator
\begin{gather*}
L_n v(s,z,a) = \,\Bigg(c_n(s,z,a)+  e^{-r_{t_{n}}\,\Delta t_{n+1}} \hspace{-0.3cm}\sum_{i \in Z^{(n+1)}} \hspace{-0.3cm}\Big( p^{(n)}_i(z) \, v\big(s-a,z^{(n+1)}_i\big) \Big)\Bigg)
\end{gather*}
for each function $v:E\rightarrow \IR_{+}$ and $0\le n < N$. By introducing $M(E):=\{v:E\rightarrow \IR_{+}\}$ the operator $L_n$ maps a function from $M(E)$, which takes two arguments $(s,z)$, to a function on $D_n$, which takes three arguments $(s,z,a)$.

\begin{theorem}

\begin{itemize}
 \item[a)] $V_n\in{M}(E)$, $V_N = 0$ and the sequence $(V_n)_{0\le n\le N}$ satisfies the Bellman equation, i.e.
\begin{align}
V_n(s,z) = \hspace{-0.3cm}\min_{a\in D_n(s,z)} L_nV_{n+1}(s,z,a)
\label{456}
\end{align}
holds for $n=0,1,\ldots,N-1$.
\item[b)] For each $n=0,\ldots,N-1$ there exists a decision rule $f_n^{*}$ such that
\begin{align*}
&V_n(s,z) = L_nV_{n+1}(s,z,f_n^{*}(s,z))
\end{align*}
holds. Such a decision rule is called a minimizer and the sequence $(f_{0}^{*},f_{1}^{*},...,f_{N-1}^{*})$ of minimizers is an optimal Markovian policy.
\end{itemize}
\label{Bellman}
\end{theorem}

\begin{remark}
The minimizer $f_n^{*}(s,z)$ gives the solution of the minimization problem in (\ref{456}) which depends on the actual state $(s,z)$.
\end{remark}

Theorem \ref{Bellman} yields that the value function $V_N$ equals $0$. By using the Bellman equation (\ref{456}) we can compute the value function $V_{N-1}$. By going on in this manner we finally compute the value function $V_0$ which is the solution of the MDP. Casually, we get an optimal policy by computing a minimizer in each step, i.e.\ the policy $\pi^{*}=(f^{*}_0,f^{*}_1,...,f^{*}_{N-1})$, where $ f^{*}_n=f_n^{*}(s,z)$ minimizes $a \mapsto L_n V_{n+1}(s,z,a)$ for all $(s,z)\in E$, is an optimal one. Since we are going backwards in time this algorithm is called {\it Backward Induction Algorithm}. Putting this algorithm on a formal basis we get:

\begin{enumerate}
\item $V_N(s,z)=0 \quad \forall (s,z)\in E$. Set $n:=N$.
\item Set $n:=n-1$ and compute for all $(s,z)\in E$
\begin{align*}
&V_{n}(s,z) =  \hspace{-0.3cm}\min_{a\in D_n(s,z)} L_{n}V_{n+1}(s,z,a)
\end{align*}
as well as a minimizer $f_n^{*}=f_n^{*}(s,z)$.
\item If $n=0$, then the value function $V_0$ is computed and an optimal Markovian policy is given by $\pi^{*}=(f_0,f_1,...,f_{N-1})$. Otherwise go to step 2.
\end{enumerate}

At the end of the Backward Induction Algorithm we obtain the value $V_0(1,\lambda_0)$, which is precisely the fair value of the bond with optional sinking feature.

\begin{remark}
\begin{itemize}
\item[a)] To include more stochastic drivers, e.g., stochastic interest rates, one can easily extend the state space $E$ of the MDP. For instance, assuming $r_t$ to be stochastic we end up with the new state space $E' := S \times Z \times \tilde{Z} $ where $\tilde{Z}$ describes the state of the short rate $r_t$. Mathematically, this causes no difficulty at all. However, numerically we then have to work with a tree approximation for the bivariate stochastic process $(r,\lambda)$, which is considerably more involved. Moreover, in particular in the high-yield sector it is quite typical that the level of the default intensity $\lambda$ exceeds the level of the interest rate $r$ by far, so that small fluctuations in the latter do not affect the pricing considerably, which allows to resort to the one-factor simplification, at least approximately. 
\item[b)] In the simple special case when $\lambda$ is deterministic, the above algorithm boils down to a deterministic dynamic program. It may be used in order to compute a so-called Z-spread for bonds with optional sinking feature. The classical Z-spread is only defined for vanilla coupon bonds. Given the discounting curve $\{r_t\}$ and a bond's market price, the bond's Z-spread is defined as the unique constant spread $z$ such that the market price is explained by the default-free pricing model with discounting curve $\{r_t+z\}$. Equivalently, within our defaultable setup this correpsonds to a zero recovery assumption $R=0$ and the default intensity being chosen constantly as $\lambda \equiv z$, see \cite{pedersen06} for a detailed explanation. In particular, when the bond has no optional sinking feature but is callable, then the determinsitic program boils down to finding the minimum over (at most) $N$ bond values: for each admissible call time point we compute the respective bond value by discounting the
respective cash flows, and the callable bond price is then the minimum. This simplified, deterministic pricing approach for callable bonds is called the "Worst Ansatz" in the marketplace, e.g.\ available via Bloomberg on the screen YAS $<$GO$>$.
\end{itemize}
\end{remark}

\section{Numeric example} \label{sec_ex}
As a concrete example we apply the following special case of a credit-equity model proposed in \cite{carr06}. All objects to appear in the sequel are formally defined on a probability space $(\Omega,\F,\IP)$ supporting a Brownian motion $\{W_t\}_{t \geq 0}$ and an independent exponential random variable $\epsilon$ with unit mean. The default intensity $\lambda$ is given by
\begin{gather*}
\lambda_t:=\lambda_0\,\Big( \frac{Z_t}{Z_0} \Big)^{2\,\beta},\quad t \geq 0,
\end{gather*}
where the process $Z=\{Z_t\}_{t \geq 0}$ is a diffusion process satisfying the stochastic differential equation
\begin{gather*}
dZ_t = Z_t\,\Big( \lambda_0\,\Big( \frac{Z_t}{Z_0} \Big)^{2\,\beta}\,dt+\sigma\,Z_t^{\beta}\,dW_t\Big),
\end{gather*}
for model parameters $Z_0>0,\lambda_0>0,\sigma>0,\beta<0$. In theory, the process $Z$ might diffuse to zero. However, it is shown in \cite{carr06} that the default time $\tau$, defined via the canonical construction (\ref{canonical}) happens almost surely before $Z$ hits zero.
This is due to the fact that $\lambda_t$ approaches infinity very quick as $Z_t$ tends to zero. Hence, $\lambda_t>0$ for all $t<\tau$ almost surely, which is sufficient for practical needs. The motivation for this modeling setup is a link between credit and equity of the same company:
The stock price of the company is modeled as $S_t=Z_t\,1_{\{\tau>t\}}$, i.e.\ as the diffusion process $Z$ until default happens, and zero after default. It can be shown that $\{S_t\}_{t \geq 0}$ is a martingale, hence this model can be used for pricing stock derivatives (depending on $S=\{S_t\}$) and credit derivatives (depending on $\lambda=\{\lambda_t\}$) jointly. This allows, for instance, to extract information about the creditworthiness of the company from equity option data.
\par
For a pre-determined time grid $0=t_0<t_1<t_2<\ldots<t_N = T$ we discretize the default intensity process by a finite-state process $\{\lambda_{t_n}\}_{n =0,\ldots,N}$. For this, we use the trinomial tree described in Appendix F of \cite{brigo01} as an approximation of the diffusion process $Z$.
Since the diffusion process $Z$ has state-dependent volatility, this approximation cannot be applied directly, but $Z$ has to be transformed first to a Bessel process (with constant volatility term) as described in \cite[Proposition 5.1]{carr06}. The desired tree approximation for $Z$ is finally obtained via backtransforming the Bessel tree node-by-node, making use of Slutzky's theorem\footnote{I.e.\ if the tree converges weakly to the Bessel process, then the transformed tree converges weakly to the process $Z$.}.
In order to incorporate the possibility of a default event we add the state $\infty$ to the possible values of $\lambda_{t_n}$ for all $n\geq 1$. At each time step $t_n$ the probability of $\lambda_{t_{n+1}}$ becoming $\infty$ is set to $1-\exp(-\lambda_{t_n}\,\Delta t_{n+1})$. Moreover, all node probabilities of the tree for $Z$ at time $t_{n+1}$ are multiplied by $\exp(-\lambda_{t_n}\,\Delta t_{n+1})$, so that all probabilities sum up to one.
\par
We illustrate our approach by pricing the aforementioned bond issued by Westvaco Corporation with optional sinking feature. All data have been retrieved via Bloomberg and the programming has been carried out in MATLAB on a standard PC. The interest rate discounting curve $\{r_t\}_{t \geq 0}$ has been bootstrapped from EONIA swap rates, US dollar interest rate swap rates and cross currency basis swap spreads, along the methods described in \cite{hagan06,fujii10}. We set the recovery rate to $R=0.4$, which is a standard assumption for senior unsecured corporate debt. Observing the stock price $S_0=Z_0$, we calibrate the remaining model parameters $(\sigma,\lambda,\beta)$ to observed quotes for credit default swaps (CDS) referencing on Westvaco. This calibration is accomplished by defining a grid for the three parameters and resorting to the built-in MATLAB function \texttt{fminsearch}, applied to an error functional which compares model and market prices and penalizes parameter choices outside a reasonable
domain. This gives us the fitted parameters $\beta = -0.6$, $\sigma=2.8199$ and $\lambda = 0.004$.
In the final analysis the JDCEV model explains the market quotes quite well since the model CDS prices lie between the market CDS bid and market CDS ask prices as we have checked via Bloomberg.

With these parameters specified, the {\it Backward Induction Algorithm} is now implemented in order to price the bond with optional sinking feature in concern. For the specified bond, the issuer has the right to redeem $10 \%$ of the outstanding nominal every year, but may also opt for a smaller redemption of only $5 \%$. In the past, a certain amount of the initial outstanding nominal has already been redeemed, so that only a fraction of $\alpha \%$ of the initial nominal are currently outstanding (and $\alpha$ is an observable multiple of $5$). Assuming we invest one dollar into the bond now, if the issuer chooses to redeem $10 \%$ of the initial outstanding nominal, this  results in a redemption of $10/\alpha$ dollar for us, whereas a choice of only $5 \%$ of the initially outstanding nominal results for us in a redemption of $5/\alpha$ dollar. This choice is valid once a year, and at these time points $t_n$ the admissible action set is therefore given by $D_n(s,z):=\{5/\alpha,10/\alpha\} \cap [0,s]$,
where $s \in S:=\{0,5/\alpha,10/\alpha,15/\alpha,\ldots,(\alpha-5)/\alpha,1\}$. At all other time points $t_n$ in the grid we set $D_n(s,z):=\{0\}$,
meaning that no redemption takes place at $t_n$. The pricing algorithm took $1.57$ seconds on a standard PC with $N=403$ grid points. The bond price computed was $115.30 \%$, which is well in line with the quoted bid-ask spreads on Bloomberg, or alternatively on public exchange websites. If the issuer was not allowed to choose between $10 \%$ or $5 \%$ redemption but instead was always forced to redeem $10 \%$ ($5 \%$), the respective bond value obtained would be $116.04 \%$ ($125.24 \%$). Hence, if the issuer had to decide immediately between these two extreme redemption schedules, she would opt for the maximal redemption amount each time. Having the additional option to switch between $5 \%$ and $10 \%$ redemption during the lifetime of the bond -- in comparison to being forced to always redeem the maximal amount of $10 \%$ -- is worth $0.74 \%= 116.04 \% - 115.30 \%$ to the issuer.

\section{Conclusion} \label{sec_con}
It was shown how to compute the price of a bond with optional sinking feature in a one-factor model for the issuer's default intensity. The approach is based on Markov Decision Processes and implemented via the Backward Induction Algorithm. We illustrated the applicability of the presented ansatz using the JDCEV model of \cite{carr06}, and included a real-world example.


\end{document}